# Revisiting the transient coarsening kinetics: a new framework in the Lifshitz-Slyozov-Wagner space

Yue Li[1], Zhijun Wang[1]*, Xianghan Gao[2], Yujian Wang[1], Junjie Li[1], Jincheng Wang[1]

[1] State Key Laboratory of Solidification Processing, Northwestern Polytechnical University, Xi'an 710072, China

[2] Department of Georesources and Materials Engineering, RWTH Aachen University, 52056 Aachen, Germany

**Abstract**

Phase coarsening is a fundamental process of microstructure evolution in multiphase materials. A thorough understanding of its kinetics is of great significance for material processing and performance. Generally, coarsening can be divided into the transient stage and the steady stage. Compared with steady coarsening kinetics, the current understanding of transient coarsening is rather limited and contradictory. In the present work, a new framework in the dimensionless Lifshitz-Slyozov-Wagner space is developed to study transient coarsening kinetics co-controlled by interface migration/reaction and matrix diffusion, where the dynamic equation for individual particles is derived from the thermodynamic extremal principle. The effects of initial particle size distribution and volume fractions are revealed. Firstly, different from the conventional viewpoints, the time for transient coarsening is found to change non-monotonically with the width and tail length of the initial distribution. The ultralong transient coarsening can occur for systems initiated from both the 'wide & long tail' and 'narrow & short tail' initial distributions, offering a new understanding of microstructure stability. Furthermore, the theoretical explanation of the 'quasi-steady' distributions in the ultralong transient stage interestingly coincides with Brown's unsolved puzzle about the unique steady state for decades [LC Brown. Acta Metall 1989;37:71]. Finally, an increase in volume fraction is shown to shorten the single mechanism-dominated transient stage, but delay the transition from the interface-dominated state to the diffusion-controlled stage. These findings not only offer new insights for the previous observations in microgravity experiments [VA Snyder, J Alkemper, PW Voorhees, Acta Mater 2001;23:699], but also theoretically describe the limiting coarsening behaviors at ultrahigh volume fractions [H Yan, KG Wang, ME Glicksman. Acta Mater 2022;117964].

* Corresponding author. zhjwang@nwpu.edu.cn

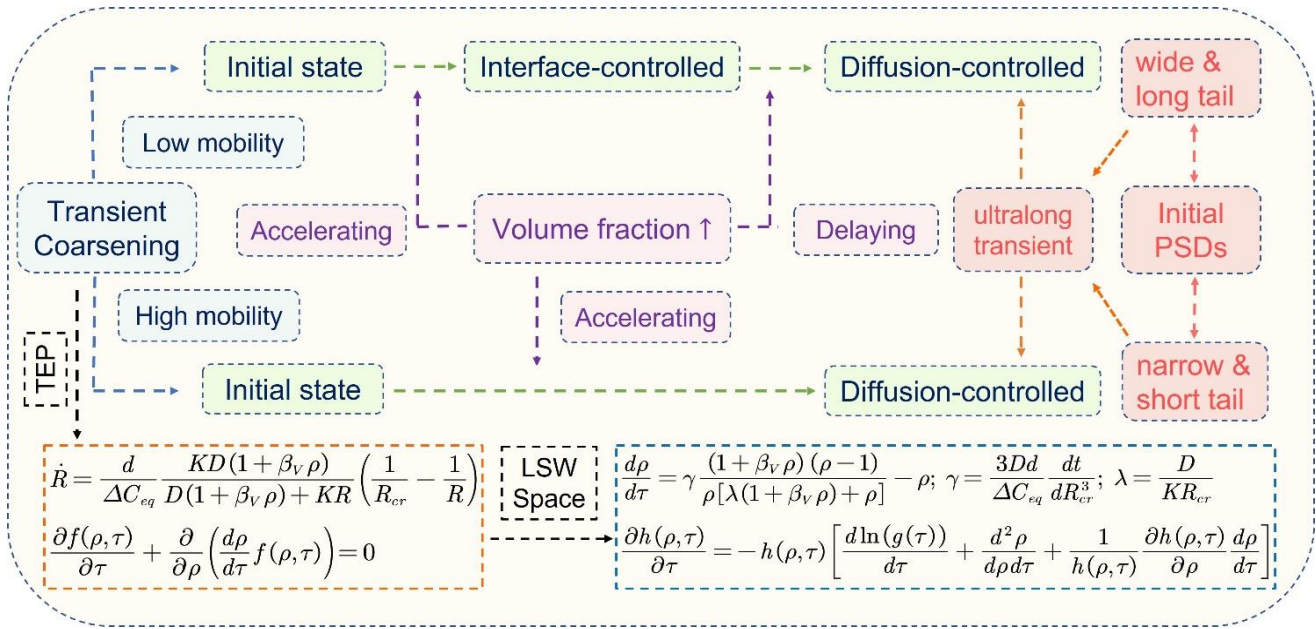

Graphical Abstract for '**Revisiting the transient coarsening kinetics: a new framework in the Lifshitz-Slyozov-Wagner space**'.

# 1. Introduction

Phase coarsening is a fundamental microstructure evolution process in multiphase materials, driven by a spontaneous decrease in the total interfacial energy. With larger particles growing at the expense of smaller ones, phase coarsening can result in an increase in average particle size and a corresponding decrease in particle number density, which will significantly affect materials' properties and performances. For instance, Wang and his coworkers **[1-4]** have systematically studied the influences of phase coarsening (average particle size and particle size distribution) on mechanical properties of materials, e.g., hardness, fracture strength and fatigue resistance. Thus, a thorough understanding of the phase coarsening kinetics is of great fundamental and practical interest to better tailor materials to desired performances.

In general, phase coarsening can be divided into two regimes: the steady stage (or called the long-time regime) and the transient stage before it. In the steady stage, there are a constant average particle growth rate and a self-similar particle size distribution (PSD). Whereas, both average growth rate and PSD are evolved in the transient stage. Starting from the seminal mean-field model by Lifshitz and Slyozov **[5]**, and Wagner **[6]**, numerous works have been established to describe the steady coarsening kinetics, especially for the finite volume fraction effects on the shape of steady PSDs and the average growth rate **[7, 8]**. However, compared with the mature understanding of the steady stage, the current understanding of the transient coarsening kinetics is rather limited and contradictory.

The scientific interests in transient coarsening began with a series of microgravity-based Pb-Sn experiments by Voorhees and co-workers **[9-12]**, where they found that the measured data deviated from theoretical predictions of the steady stage, even though the average radius had changed by a factor of three. In contrast, there was a good agreement between the measured data and numerical simulations of the transient coarsening by the multiparticle diffusion approach **[9-11]** and the mean-field approach **[9]**. These works showed the significance of the transient stage, and the shape of initial distributions and finite volume fractions were found to affect the transient kinetics significantly **[13]**.

Generally, width and tail length are two common factors characterizing the shape of the initial distribution, but the current understanding of these factors is highly controversial. For the width, Chen and Voorhees's mean-field simulations [13] found that the transient stage for initial wide distributions was shorter than that for the narrow ones. However, Voorhees and Glicksman [14] (multiparticle diffusion approach) and Smet et al. [15] (mean-field approach) have reached an exact opposite conclusion where the transient stage for initial wide distributions was longer. Differently, the results of the mean-field simulation by Enomoto et al. [16] and phase-field simulation of some current authors [17] have found that there was no observable effect of the initial width on the transient length. In addition, Fang et al. [18] found that instantaneous coarsening rate (in transient stage) with wider initial distribution is higher, which is also consistent with Nandal et al.'s experimental results [19]. As for the effect of tail length, the phase-field simulations of some current authors [17] have shown that the coarsening system with initial PSD of a longer tailer usually shows a longer transient stage, while the influence of width is insignificant. In addition, Fang et al. [18] found that instantaneous coarsening (in transient stage) with longer initial tail is faster.

About effects of initial distribution, there exists another long-standing problem without a definitive conclusion, i.e., whether there only exists a unique attractor state for the steady stage. This problem was firstly proposed by Brown with mean-field [20, 21] and multiparticle diffusion simulations [22], where he found that there existed some other 'self-similar' distributions besides the solutions from the LSW asymptotic analysis, starting from some specific initial distributions. And a similar conclusion was also found in grain-growth systems [23, 24]. Then, the existence of multiple stable distributions and volume conservation criteria were discussed by Hillert et al. [25, 26], Brown [27-29], Coughlan and Fortes [30], and Hoyt [31]. Finally, it was agreed that those extra 'self-similar' distributions were not truly stable, whose remarkable stability (depending on the shape of initial distributions) may be attributed to the numerical procedure, but detailed origin was not revealed yet. Therefore, it is important to clarify the effects of initial distributions on transient coarsening.

As for the effect of volume fractions on transient coarsening kinetics, the existing problem was the great disagreement between experimental results and numerical simulations. Snyder et al.'s microgravity-based Pb-Sn experiments **[11]** have found that the transient stage for higher volume fractions is longer. Whereas the Chen and Voorhees's mean-field simulations **[13]** have reached an exact opposite conclusion, which is also consistent with the phase-field simulations of some current authors **[17]**. One important thing should be realized is that all these simulations were for the matrix diffusion-controlled coarsening kinetics. But in fact, the particle growth or shrinkage usually goes through two continuous processes, i.e., matrix diffusion and interface migration (or interface reaction **[32, 33]**). When the latter is significantly slower, the coarsening kinetics is mainly determined by the interface process.

For coarsening co-controlled by interface and matrix diffusion, White **[34, 35]** and Sun **[36]** have found that the rate-determining strength of the interface process would increase with a descent of the interfacial rate but decay as an increment of the average particle radius (with time going by). Therefore, the transition from the interface-controlled state to the diffusion-controlled steady stage can also be viewed as a 'special' transient stage. Qualitatively, with an increase in volume fraction, the rate-determining strength of the interface process will increase due to the faster matrix diffusion, and thus result in a longer transient stage for some specific coarsening systems, e.g., the solid-liquid (Pb-Sn) system, where the interfacial rate of the solid particles is much smaller than the matrix diffusion. Therefore, it is necessary to quantitatively reveal the effects of volume fraction on this 'special' transient coarsening kinetics, which may offer a reasonable explanation for the previous disagreement between experiments and simulations.

Another interesting problem related to this 'special' transient stage is the limiting behaviors of coarsening kinetics at ultrahigh-volume fractions. In Wang et al.'s phase-field simulations for coarsening at ultrahigh-volume fraction regime **[37-40]**, they found an interface-controlled (or grain growth) like particle size distribution and an intermediate scaling growth exponent between 3 (diffusion controlled) and 2 (interface controlled), which decreased with an increase in volume fraction **[40]**. This result suggests a mixed mode coarsening at ultrahigh volume fractions, and the rate-determining strength of the interface mechanism increases with volume fractions. Revealing effects of volume fraction on transient coarsening may provide a more detailed description on this interesting scientific topic.

To systematically revisit the previous controversies about transient coarsening kinetics, this article develops a new numerical framework for studying transient coarsening co-controlled by interface and matrix diffusion, whose dynamic equation for individual particles is deduced from the thermodynamic extremal principle. And the organization of this paper follows as: in **Section 2**, there is a short review of the existing coarsening theory for better comparison with this work. In **Section 3**, we derive the dynamic equations of coarsening co-controlled by interface and matrix diffusion and establish the present numerical framework in the LSW space (dimensionless). In **Section 4**, there are numerical details and verifications of the present simulations. And **Section 5** shows simulation results of different transient coarsening systems and presents a systematic analysis of the effects of initial PSD shape and finite volume fractions. Finally, conclusions are drawn in **Section 6**.

## 2. A short review of coarsening theory

In the seminal mean-field model by Lifshitz and Slyozov **[5]**, and Wagner **[6]** (referred to as the 'LSW' theory) for a vanishingly small volume fraction of the precipitates ($\phi_V \to 0$), they derived the dynamic equation of individual particles by neglecting the interparticle diffusional interaction

$$\left(\dot{R}\right)_{LSW} = \frac{D}{\Delta C_{eq}} \frac{1}{R} (C_\infty - C_R) = \frac{D}{\Delta C_{eq}} \frac{1}{R} \left(\frac{d}{R_{cr}} - \frac{d}{R}\right), \tag{1}$$

where $D$ is the matrix diffusion constant, $d$ is the capillary length, $\Delta C_{eq}$ denotes the equilibrium concentration difference between precipitates and matrix, $C_\infty$ and $C_R$ represent the concentrations in equilibrium with individual particle's radius $R$ and critical particle radius $R_{cr}$, respectively.

Combining **Eq. (1)** with the continuity equation of distribution function $F(R,t)$ in the radius space

$$\frac{\partial F(R,t)}{\partial t} + \frac{\partial}{\partial R}\left(\dot{R}F(R,t)\right) = 0, \tag{2}$$

the steady dimensionless distribution $h(\rho)$ and growth equation of average radius $\langle R \rangle$ yielded as

$$h(\rho)_{LSW} = \begin{cases} \dfrac{3^4 e \rho^2}{2^{5/3}(3+\rho)^{7/3}(3/2-\rho)^{11/3}} \exp\left(\dfrac{3/2}{\rho-3/2}\right), & \rho \leq \dfrac{3}{2} \\ 0, & \rho > \dfrac{3}{2} \end{cases} \tag{3}$$

$$\langle R \rangle^3 - \langle R_0 \rangle^3 = \frac{4Dd}{9\Delta C_{eq}}(t - t_0). \tag{4}$$

In later years, numerous theories have been further developed to describe coarsening kinetics at finite volume fractions, which can be classified into two categories: (i) effective medium theories from average sphere models (or called *ad hoc* assumptions) **[41-45]** or diffusion screening concepts **[46-50]**; (ii) statistical mechanical models considering the particle interactions by means of surrounding diffusive fluxes **[51-57]**. Among these works, we emphasize an analytical model for $\phi_V = 0 \to 1$ developed by Streitenberger **[45]**, whose change rate of individual particles can be represented by

$$\dot{R} = \frac{D}{\Delta C_{eq}} \frac{1}{R}(1 + \beta_V \rho)\left(\frac{d}{R_{cr}} - \frac{d}{R}\right); \tag{5}$$

$$\beta_V = \frac{4\left(\phi_V \dfrac{\mu_1}{\mu_3}\right)^{1/2}}{\left[1 - \left(\phi_V \dfrac{\mu_1}{\mu_3}\right)^{1/2}\right]\left[3 + \left(\phi_V \dfrac{\mu_1}{\mu_3}\right)^{1/2}\right]}, \tag{6}$$

where the finite volume fraction effects are represented by the kinetic parameter $\beta_V$, and $\mu_k = \int \rho^k h(\rho) d\rho$ is the 'kth' moment of the $h(\rho)$. **Eqs. (5)** and **(6)** can provide elegant analytical solutions for the steady PSDs ($h_s(\rho)$) and growth equation of $\langle R \rangle$, yielding as

$$h_s(\rho) = \begin{cases} \dfrac{Ab^p \rho_{\max}^q \exp(r)\rho^2}{(b+\rho)^{1+p} \cdot (\rho_{\max}-\rho)^{1+p}} \exp\left(\dfrac{r\rho_{\max}}{\rho-\rho_{\max}}\right), & \rho \leqslant \rho_{\max} \\ 0, & \rho > \rho_{\max} \end{cases}$$

$$\rho_{\max} = \frac{3}{2} + \frac{\phi_V}{2}, \quad b = \frac{(2-\rho_{\max})\rho_{\max}}{(\rho_{\max}-1)^2}, \quad p = \frac{Ab^2}{(b+\rho_{\max})^2}, \tag{7}$$

$$q = \frac{A\rho_{\max}(2b+\rho_{\max})}{(b+\rho_{\max})^2}, \quad r = \frac{A\rho_{\max}}{b+\rho_{\max}},$$

$$\langle R \rangle^3 - \langle R_0 \rangle^3 = \mu_1^3 \frac{4Dd}{9\Delta C_{eq}} \frac{(1+\phi_V^{1/2})^2}{(1-\phi_V^{1/2})(1+\phi_V^{1/2}/3)^2}(t-t_0). \tag{8}$$

where $\rho_{\max}$ is the maximum relative radius and $A$ is a normalization constant. For volume fractions ranging from $\phi_V = 0$ to $\phi_V = 0.3$, **Eq. (6)** is very close to that derived from diffusion screening theory **[49, 50]** or statistical theory **[51-53]**, yielding as

$$\lim_{\phi_V \to 0} \beta_V \approx \sqrt{3\phi_V \frac{\mu_1}{\mu_3}}. \tag{9}$$

The validity of **Eq. (9)** has been systematically tested by Wang et al. through comparing their diffusion screening theory **[49, 50]** with multiparticle diffusion simulations **[50, 58]** and ground-based experiments of Al-Li alloy system **[59]**. As for volume fractions $\phi_V > 0.3$, **Eq. (6)** also has a good agreement with Kim et al.'s phase-field results **[60]** and Wang et al.'s very recent integrated work of phase-field simulations and microgravity-based Pb-Sn experiments **[61, 62]**.

As for the interface-controlled coarsening, Wagner **[6]** gave the change rate of particles

$$\left(\dot{R}\right)_{Interface} = \frac{K}{\Delta C_{eq}}(C_\infty - C_R) = \frac{K}{\Delta C_{eq}}\left(\frac{d}{R_{cr}} - \frac{d}{R}\right), \tag{10}$$

where $K$ is an interfacial rate constant, whose detailed expression is given in **Chapter 4** of **Ref. [8]**. The steady distribution $h(\rho)$ and growth equation of $\langle R \rangle$ were solved to be

$$h(\rho)_{Interface} = \begin{cases} \dfrac{24\rho}{(2-\rho)^5} \exp\left(\dfrac{3\rho}{\rho-2}\right), & \rho \leqslant 2 \\ 0, & \rho > 2 \end{cases} \tag{11}$$

$$\langle R \rangle^2 - \langle R_0 \rangle^2 = \frac{Kd}{2\Delta C_{eq}}(t - t_0). \tag{12}$$

And **Eq. (11)** is also consistent with Hillert's distribution for normal grain growth **[63]**.

## 3. Model derivations

### 3.1. Derivations of growth rate $\dot{R}$

As one of the approved effective methods for modeling microstructure evolution **[64]**, the thermodynamic extremal principle (TEP) is adopted here for simply considering the influence of the finite interface mobility on coarsening. Under a sphere approximation of precipitates, the total interfacial energy of the system can be represented by

$$G = \int_0^{R_{\max}} \{4\pi\sigma R^2 F(R,t)\} dR, \tag{13}$$

where $\sigma$ is the specific energy of the matrix/precipitate interface, and $R_{\max}$ is the maximum particle radius at time $t$. Thus, the change rate of the free energy yields as

$$\dot{G} = \int_0^{R_{\max}} \left\{4\pi\sigma R^2 \frac{\partial F(R,t)}{\partial t}\right\} dR + 4\pi\sigma R_{\max}^2 F(R_{\max},t) \frac{dR_{\max}}{dt}, \tag{14}$$

where the second term of **Eq. (14)** equals 0 since that $F(R_{\max},t) = 0$. Then, inserting the continuity equation of $F(R,t)$ **[Eq. (2)]** into **Eq. (14)** and applying that $F(0,t) = F(R_{\max},t) = 0$, the final expression of $\dot{G}$ can be transformed by partial integral, yielding as

$$\dot{G} = \int_0^{R_{\max}} \{8\pi\sigma R \dot{R} F(R,t)\} dR. \tag{15}$$

In the case of coarsening co-controlled by interface migration and matrix diffusion, the total dissipation includes two components **[65, 66]**, i.e., $Q_M$ associated with the interfacial migration and $Q_D$ associated with matrix diffusion. The $Q_M$ in 3-D configuration can be denoted by **[65, 66]**

$$Q_M = \int_0^{R_{\max}} \left\{ \frac{4\pi R^2 \dot{R}^2}{M} F(R,t) \right\} dR, \tag{16}$$

where $M$ is the mobility of interface. As for $Q_D$, we assume the same diffusion zone as Svoboda and Fischer's work ($r: R \to U$) **[44]**,

$$Q_D = \int_0^{R_{\max}} \left\{ R_g T \left[ \int_R^U \left( 4\pi r^2 \frac{j_R^2(r)}{C^* D} \right) dr \right] F(R,t) \right\} dR, \tag{17}$$

where $R_g$ is the gas constant, $C^*$ is the equilibrium concentration of the matrix, and $j_R(r)$ is the radial diffusive flux in the matrix surrounding the particle of radius $R$

$$j_R(r) \approx \Delta C_{eq} \frac{R^2}{r^2} \dot{R}, \ R \leqslant r \leqslant U. \tag{18}$$

In order to reproduce the Streitenberger's analytical model for $\phi_V = 0 \to 1$ **[45]**, Ardell concept **[41]** for outer boundary of the diffusion zone ($U = R + R_{cr}/\beta_V$) is adopted in **Eq. (16) [44]**, which combined with **Eq. (18)** transforms **Eq. (17)** into

$$Q_D = R_g T \frac{\Delta C_{eq}^2}{C^* D} \int_0^{R_{\max}} \left\{ 4\pi R^4 \left( \frac{1}{R} - \frac{1}{R + R_{cr}/\beta_V} \right) \dot{R}^2 F(R,t) \right\} dR. \tag{19}$$

Now, assuming the total volume of matrix and precipitates as 1, the change rate of precipitates' volume fraction yields as

$$\phi_V = \int_0^{R_{\max}} \left\{ \frac{4\pi R^3}{3} F(R,t) \right\} dR \to \frac{d\phi_V}{dt} = \int_0^{R_{\max}} \left\{ 4\pi R^2 \dot{R} F(R,t) \right\} dR. \tag{20}$$

Although one can apply the volume conservation ($d\phi_V/dt$) as a simple approximation, it is better to use the mass conservation **[68]**, especially for the transient coarsening stage **[13]**, due to significant changes of volume fractions at an early stage of coarsening **[60]**. As approximated by Marqusee and Ross **[67]**, the mass conservation can be simply estimated by

$$\phi_V \approx \phi_{V,\,eq} - \frac{d}{\Delta C_{eq} R_{cr}} \to \frac{d\phi_V}{dt} \approx \frac{d}{\Delta C_{eq} R_{cr}^2} \frac{dR_{cr}}{dt}. \tag{21}$$

Therefore, the thermodynamic variations in TEP [64] can be represented by

$$\frac{\delta}{\delta \dot{R}} \left\{ \dot{G} + \frac{Q_D + Q_M}{2} + L\left[ \frac{d}{\Delta C_{eq} R_{cr}^2} \frac{dR_{cr}}{dt} - \int_0^{R_{max}} \left\{ 4\pi R^2 \dot{R} F(R,t) \right\} dR \right] \right\} = 0, \quad (22)$$

where $L$ is the Lagrange multiplier for mass conservation, and the solved $\dot{R}$ yields as

$$\dot{R} = \frac{MD(1+\beta_V \rho)}{D(1+\beta_V \rho) + \frac{R_g T \Delta C_{eq}^2}{C^*} MR} \left( \frac{2\sigma}{R_{cr}} - \frac{2\sigma}{R} \right). \quad (23)$$

To compare with **Eqs. (5)** and **(10)**, defining that $K = 2M\sigma\Delta C_{eq}/d$ and $d = 2\sigma C^*/R_g T \Delta C_{eq}$, **Eq. (23)** transforms into

$$\dot{R} = \frac{d}{\Delta C_{eq}} \frac{KD(1+\beta_V \rho)}{D(1+\beta_V \rho) + KR} \left( \frac{1}{R_{cr}} - \frac{1}{R} \right). \quad (24)$$

Now, introducing the dimensionless radius and time as $\rho = R/R_{cr}$ and $\tau = \ln[R_{cr}/R_{cr}(0)]$, the dimensionless expression of **Eq. (24)** yields as

$$\frac{d\rho}{d\tau} = \gamma \frac{(1+\beta_V \rho)(\rho - 1)}{\rho[\lambda(1+\beta_V \rho) + \rho]} - \rho;$$

$$\gamma = \frac{3Dd}{\Delta C_{eq}} \frac{dt}{dR_{cr}^3}; \quad \lambda = \frac{D}{KR_{cr}}. \quad (25)$$

When $K$ or $R_{cr}$ is significantly small, i.e., $\lambda \to \infty$, the coarsening is mainly dominated by interface migration [6]. When $\lambda \to 0$, the coarsening is mainly dominated by matrix diffusion [45]. This relationship is consistent with theoretical analysis by Wang and his co-workers [40].

### 3.2. Derivations of numerical framework

With $\rho = R/R_{cr}$ and $\tau = \ln[R_{cr}/R_{cr}(0)]$, the dimensionless expression of the continuity equation reads as

$$\frac{\partial f(\rho,\tau)}{\partial \tau} + \frac{\partial}{\partial \rho} \left( \frac{d\rho}{d\tau} f(\rho,\tau) \right) = 0, \quad (26)$$

where $f(\rho,\tau) = F(R,t)(dR/d\rho) = F(R,t)/R_{cr}$. Then, a dimensionless distribution shape function $[h(\rho,\tau)]$ is defined as

$$f(\rho,\tau) = h(\rho,\tau) \cdot g(\tau), \tag{27}$$

where $g(\tau)$ denotes the particle number density

$$g(\tau) = \frac{\phi_V}{(4\pi/3)R_{cr}^3} \approx \frac{\phi_{V,eq}}{(4\pi/3)R_{cr}^3} - \frac{d}{\Delta C_{eq}(4\pi/3)R_{cr}^4}. \tag{28}$$

Thus, only $dR_{cr}/dt$ in **Eq. (25)** is unknown, which can be obtained by inserting **Eq. (24)** into the mass conservation in **Eq. (22)**, yielding as

$$\frac{d}{\Delta C_{eq} R_{cr}^2} \frac{dR_{cr}}{dt} = \int_0^{R_{\max}} \left\{ 4\pi R^2 \dot{R} F(R,t) \right\} dR$$

$$\rightarrow \frac{dR_{cr}}{dt} = \frac{3\phi_V}{R_{cr}} \int_0^{\rho_{\max}} \left\{ \left[ \frac{D(1+\beta_V\rho)\rho(\rho-1)}{\lambda(1+\beta_V\rho)+\rho} \right] h(\rho,\tau) \right\} d\rho. \tag{29}$$

Now, combining the **Eqs. (25-29)** can offer the present numerical framework for transient coarsening

$$\frac{\partial h(\rho,\tau)}{\partial \tau} = -h(\rho,\tau) \left[ \frac{d\ln(g(\tau))}{d\tau} + \frac{d^2\rho}{d\rho d\tau} + \frac{1}{h(\rho,\tau)} \frac{\partial h(\rho,\tau)}{\partial \rho} \frac{d\rho}{d\tau} \right]. \tag{30}$$

It is worthwhile to mention that the present numerical framework is similar to Chen and Voorhees's framework for transient coarsening [13] and Svoboda et al.'s recent numerical method for transient grain growth [68]. The only difference is that the time-dependent shape function $h(\rho,\tau)$ defined here is also respect to the dimensionless time $\tau = \ln(R_{cr})$ in the LSW space [5, 6], which is relatively more transparent for analyzing the mathematical characteristic of the evolving distributions, e.g., revealing the numerical origin of Brown's 'quasi-steady' distributions [20-22].

## 4. Numerical details and model verifications

### 4.1. Numerical details

In this study, four different initial PSDs, called '**super narrow & short tail**', '**narrow & short tail**' (denoted by '**super narrow**' and '**narrow**' for simplicity), '**wide & long tail**', and '**wide & short tail**', are constructed to clarify the effects of initial PSDs' width and tail length, as shown in **Fig. 1**. The parameters used in this work are as follows: equilibrium concentration difference $\Delta C_{eq} = 0.1$, capillary length $d=0.01$, real time step $\Delta t=0.1$ with $N_{step}=10^6$, initial particle radius $R_{cr}(0)=5$, and $\phi_V=0.1$, $0.4$, $0.7$ for investigating the effects of finite volume fractions. As for the competition between interface migration and matrix diffusion, the interface coefficient and diffusion coefficient are given as **Case-1** ($D=10$; $K=1000$) for diffusion-controlled mechanism and **Case-2** ($D=10$; $K=1$) for coarsening co-controlled by interface and matrix diffusion.

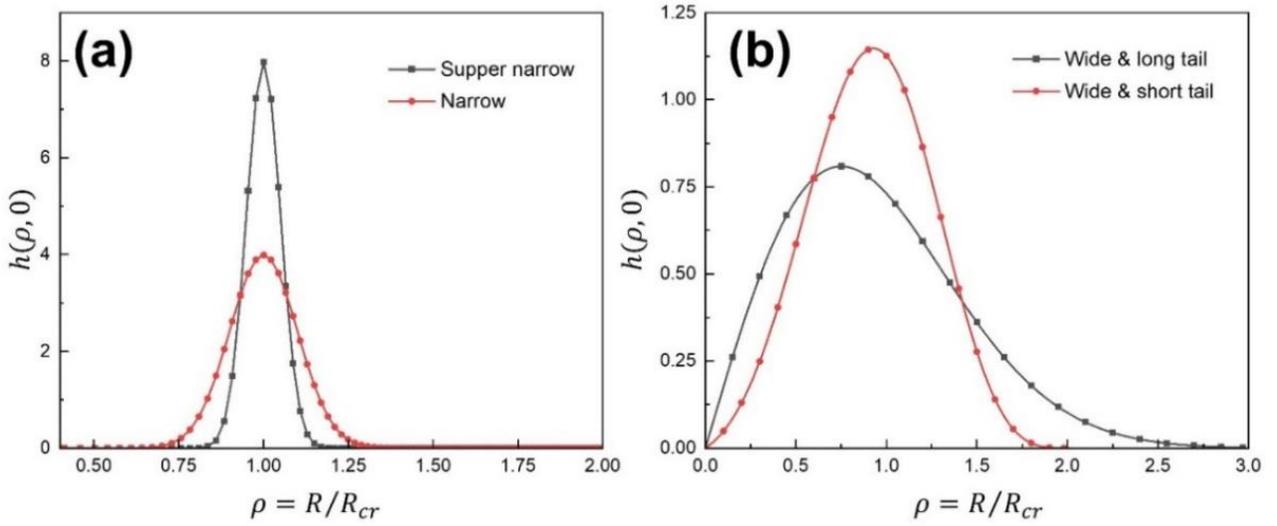

**Fig. 1.** Different initial distributions used in simulations: (a) initial PSDs; (b) wide PSDs.

### 4.2. Model verifications

To test the validity of the present numerical framework, it is necessary to compare its predictions for the steady stage with those of Streitenberger's theory **[45]**. The simulations of **Case-1** (diffusion-controlled case) with '**narrow**' initial PSD are found to approach the steady stage quickly. Although the present simulation applies the mass conservation, the predictions on steady PSDs still exhibit a good agreement with the analytical solutions in **Eq. (7)**, as shown in **Fig. 2(a)**. In addition, **Fig. 2(b)** compares the predictions on steady $dR_{cr}^3/dt$ normalized to the LSW result $dR_{cr}^3/dt = 4Dd/9\Delta C_{eq}$, where the black points denote the present numerical simulations, and the red curve represents the Streitenberger's theory **[45]** by approximating $\mu_1/\mu_3 \approx 1$ in **Eq. (6)**, yielding as

$$\frac{dR_{cr}^3}{dt}\frac{9\Delta C_{eq}}{4Dd} = \frac{\left[1+\left(\phi_V \frac{\mu_1}{\mu_3}\right)^{1/2}\right]^2}{\left[1-\left(\phi_V \frac{\mu_1}{\mu_3}\right)^{1/2}\right]\left[1+\left(\phi_V \frac{\mu_1}{\mu_3}\right)^{1/2}/3\right]^3} \approx \frac{(1+\phi_V^{1/2})^2}{(1-\phi_V^{1/2})(1+\phi_V^{1/2}/3)^3}. \tag{31}$$

Compared the black points with red curve, there exists a significant deviation between Streitenberger's theory and the present simulation for $\phi_V = 0.7$. However, if estimating $\mu_1/\mu_3$ by the simulated values of $\mu_1$ and $\mu_3$, the difference will be small, as shown by the blue points. Therefore, the present numerical framework can yield accurate predictions for the steady coarsening kinetics.

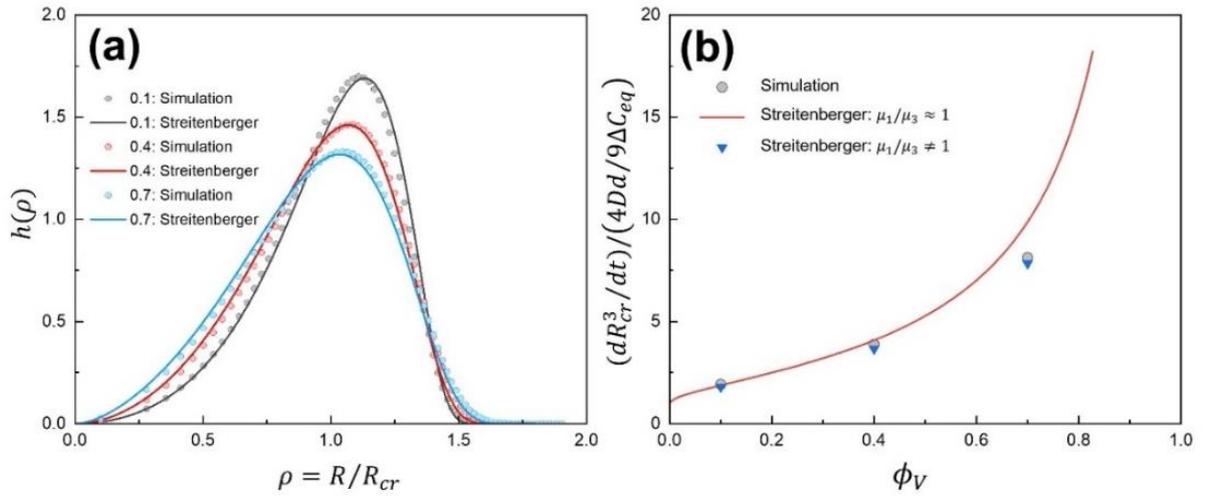

**Fig. 2.** Comparison of present simulations for **Case-1** with '**narrow**' initial PSD at the $10^6$ step with those of Streitenberger's theory [45]: (a) steady PSDs; and (b) steady $dR_{cr}^3/dt$ normalized to the LSW results.

## 5. Results and discussions

### 5.1. Effects of initial width and tail length

In order to compare with the previous works of the initial distribution's effects, the simulation results of **Case-1** (diffusion-controlled case) are presented and discussed in this part. Started from four initial PSDs, the temporal evolutions of $R_{cr}$ against $t^{1/3}$ for $\phi_V = 0.1, 0.4, 0.7$ are plotted in **Fig. 3**, where there is about a factor of $8-14$ change in $R_{cr}$. Such a long timescale can provide a credible examination of the diffusion-controlled transient coarsening kinetics.

14 / 31

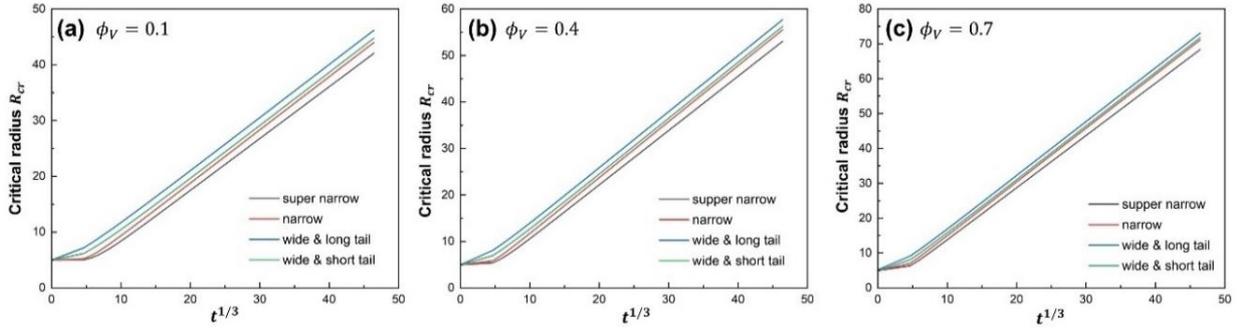

**Fig. 3.** Temporal evolutions of $R_{cr}$ for (a) $\phi_V = 0.1$, (b) $\phi_V = 0.4$, and (c) $\phi_V = 0.7$.

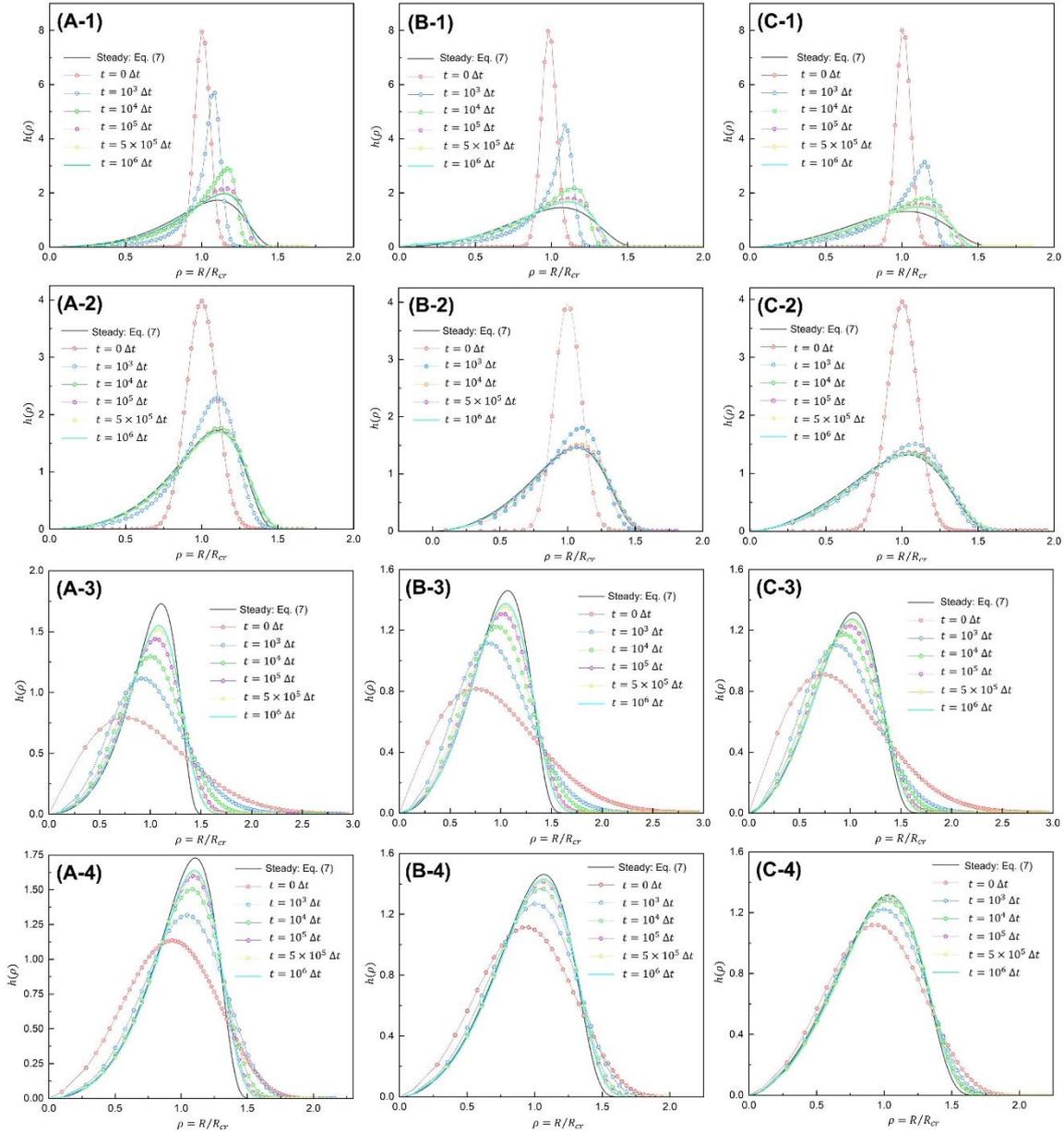

**Fig. 4.** PSD evolutions for (A) $\phi_V = 0.1$, (B) $\phi_V = 0.4$, and (C) $\phi_V = 0.7$, where (1-4) represent the initial 'super narrow', 'narrow', 'wide & long tail', and 'wide & short tail', respectively.

And the temporal evolutions of PSD for $\phi_V = 0.1$, $0.4$, $0.7$ are plotted in **Fig. 4(A), (B)** and **(C)**, where **(1-4)** represent the initial '**super narrow**', '**narrow & short tail**', '**wide & long tail**', and '**wide & short tail**', respectively. From **Fig. 4**, it can be found that there exist significant differences between the steady PSDs and those of the '**supper narrow**' and '**wide & long tail**', which indicates the existence of long transient stage. However, it is difficult to quantitatively compare the transient time length solely from the evolutions of PSD. As Li et al. **[17]** analyzed, the evolution of scaled evolution equation, i.e., **Eq. (24)**, is more proper to determine the transient time length, compared with the other factors, e.g., evolutions of PSD and PSD's standard deviations. For the diffusion-controlled case, the expression of **Eq. (24)** ($\lambda \rightarrow 0$) is only determined by the dimensionless parameter $\gamma$. Since a good representation of the steady coarsening kinetics by '**narrow**' group in **Fig. 2**, its simulated $\gamma_n$ values are applied as reference. The proximity of system to the steady can be represented by the proximity of $\gamma/\gamma_n$ to $1$.

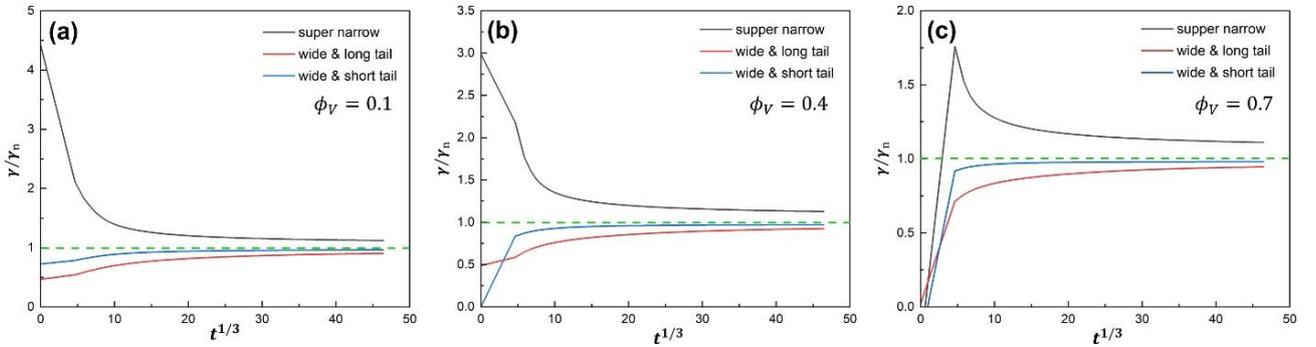

**Fig. 5.** Simulated $\gamma$ values of '**supper narrow**', '**wide & short tail**', and '**wide & long tail**' for (a) $\phi_V = 0.1$, (b) $\phi_V = 0.4$, and (c) $\phi_V = 0.7$, normalized to $\gamma$ values of the '**narrow**'.

At an early stage of transient coarsening, **Fig. 5** shows that all curves firstly undergo a drastic change, including the value of $\gamma/\gamma_n$ and slopes of curves. Then, although still away from the steady stage, the changes of $\gamma/\gamma_n$ and slopes become significantly slower. This tendency is also consistent with the evolutions of PSD, where the shape of PSD will change significantly at the beginning, while becomes 'near-similar' at the later, despite that they are still at the transient stage.

In addition, as compared in **Fig. 5**, the transient stage for '**wide**' groups show is longer than the reference group ('**narrow**'), consistent with Chen and Voorhees's conclusions of initial width's influence **[13]**. As for the '**wide**' groups, it can be found that the transient stage for '**wide & long tail**' is longer than the '**wide & short tail**', which agrees with Li et al.'s conclusions of tail length **[17]**. However, these conclusions are all contrary to the longest transient stage of the '**supper narrow**' (with **short tail**) group. In addition, if assuming all precipitates have the same radius (a line as the initial PSD), there would be no competitive advantage for any individual particles, and thus coarsening would not happen. This idealized example also supports the present simulations for '**supper narrow**'. Therefore, the previous conclusions for effects of initial width and tail length **[13, 17]** are inadequate in representing the relationship between the length of the transient stage and the shape of initial distributions. At least, from a qualitative aspect, initial distributions with both '**wide & long tail**' and '**(supper) narrow & short tail**' can show a very long transient stage. It is worthwhile to mention that the average growth rate of the different systems. As shown in **Figs. 3-5**, the instantaneous coarsening rates of the '**wide**' systems are higher than the steady coarsening rate, consistent with Fang et al.'s conclusion **[18]**. But the instantaneous rates of '**narrow**' groups are smaller, especially for the '**super narrow**', which indicates a better thermal stability from an engineering aspect

**5.2. Brown's 'quasi-steady' distributions**

As shown in **Fig. 4 and 5**, although the '**supper narrow**' and '**wide & long tail**' did not reach the steady stage, the evolutions of PSDs and scaled evolution equation [**Eq. (24)**] are very slow, which may be called as a 'quasi-steady' stage. Analogous to Brown's treatment **[20-22]**, the 'quasi' self-similar PSDs in **Fig. 4** for '**supper narrow**' and '**wide & long tail**' can be fitted by distribution functions solved by combing **Eqs. (25)** and **(26)**. In the derivation of **Eq. (7)**, the value of $\gamma$ is uniquely determined by the LSW-asymptotic analysis **[5, 6]**, yielding as

$$\frac{d\rho}{d\tau}\bigg|(\rho_{\max})=0;\ \frac{d^2\rho}{d\rho d\tau}\bigg|(\rho_{\max})=0 \to \gamma(steady)=\frac{27}{4}\frac{(1-\phi_V^{1/2})(1+\phi_V^{1/2}/3)^3}{(1+\phi_V^{1/2})^2}. \tag{32}$$

However, the $\gamma$ here is arbitrary as a tunable parameter, where the fitting results are given in **Fig. 6**. Now, it is shown that the Brown's previous simulation results **[20-22]** are reproduced, although they are not strictly self-similar.

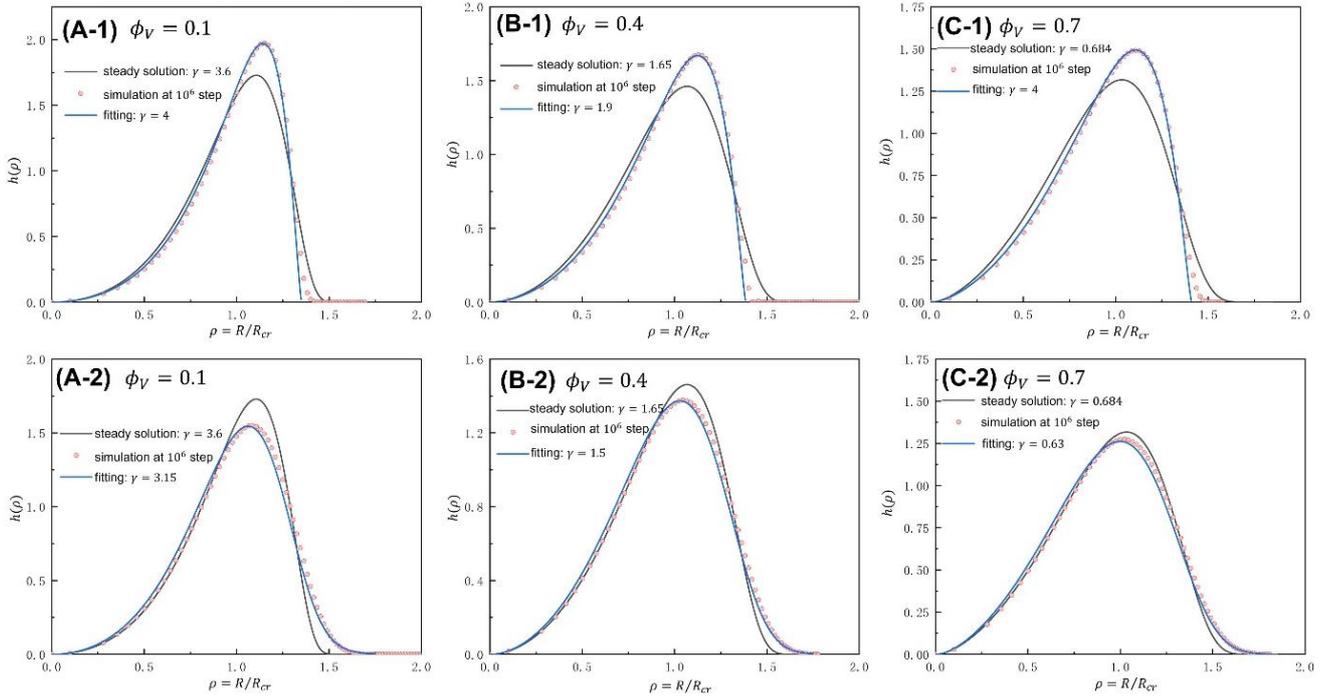

**Fig. 6.** Fitting of the 'quasi' self-similar PSDs of (1) 'super narrow' and (2) 'wide & long tail' by combining **Eqs. (25)** and **(26)** with a tunable $\gamma$.

The numerical origin of such 'quasi' self-similar PSDs can be understood from the continuity equation. For the time-independent steady PSDs $h_s(\rho)$, the dimensionless continuity equation, **Eq. (30)**, can be represented by

$$\frac{d\ln[h_s(\rho)]}{d\rho}=-\left(\frac{d\ln[g(\tau)]}{d\tau}+\frac{d^2\rho}{d\rho d\tau}\right)\frac{d\tau}{d\rho}. \tag{33}$$

Similar to **Eq. (33)**, we can define a function for transient PSDs $h(\rho,\tau)$, yielding as

$$\frac{d\ln[\bar{h}(\rho,\tau)]}{d\rho} = -\left(\frac{d\ln[g(\tau)]}{d\tau} + \frac{d^2\rho}{d\rho d\tau}\right)\frac{d\tau}{d\rho}, \tag{34}$$

where $\bar{h}(\rho,\tau)$ is an artificial assumed distribution function determined by the instantaneous scaled evolution equation [**Eq. (24)**], thus, **Eq. (30)** can be rewritten as

$$\frac{\partial\ln[h(\rho,\tau)]}{\partial\tau} = \frac{d\tau}{d\rho}\left[\frac{d\ln[\bar{h}(\rho,\tau)]}{d\rho} - \frac{\partial\ln[h(\rho,\tau)]}{\partial\rho}\right]. \tag{35}$$

Therefore, it is the instantaneous difference between $\bar{h}(\rho,\tau)$ and $h(\rho,\tau)$ that determines the 'driving force' for evolution of PSDs. When the right hand of **Eq. (35)** equals 0, it is the rigorous steady state determined by the LSW analysis **[5, 6]**. However, in the transient coarsening stage, it is possible for the evolving $\gamma$ (that determines $\bar{h}(\rho,\tau)$) to match the instantaneous distribution $h(\rho,\tau)$. It may cause a very small driving force for PSDs' evolutions, and the transient PSDs can be thought of locating in a 'quasi' self-similar regime. Then, the transient stage would be significantly long like the present '**supper narrow**' case, which can significantly deviate from the steady PSDs and coarsening rate, as shown in **Figs. 3** and **6**, even if there is a factor of $8-14$ change in $R_{cr}$. Such increment is significantly larger than the existing experimental observations **[9-12]**, which indicates the importance of understanding the ultralong transient kinetics.

Therefore, it is shown that the existence of 'quasi-steady' stage may significantly affect the overall coarsening kinetics. However, according to the present simulations, the 'quasi-steady' states perhaps occur only for some 'extreme' initial distributions, e.g., very narrow or wide with a very long tail. More efforts are still needed to reveal the evolving character of the transient PSDs and identify what kind of initial distribution can go through a significantly 'quasi-steady' stage. For instance, Streitenberger and Zöllner **[69, 70]** have developed a systematic envelop theorem for the evolving path of coarsening distributions. Although these works are aimed at describing evolution of $F(R,t)$ rather than dimensionless $h(\rho,\tau)$, the underlying insight is still of great importance for us to further explore the transient coarsening kinetics.

## 5.3. Effects of volume fractions

### 5.3.1. Shortening effect on transient stage controlled by matrix diffusion

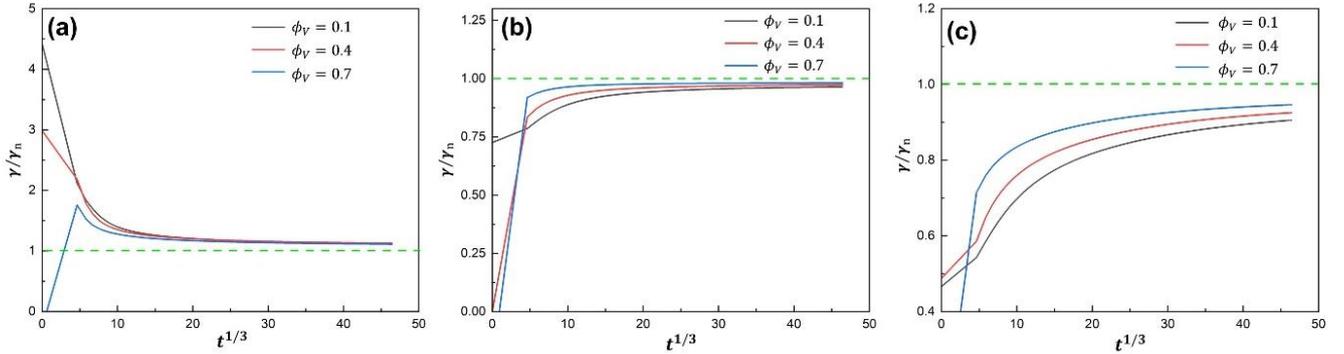

**Fig. 7.** Simulated $\gamma$ values for (a) '**supper narrow**', (b) '**wide & short tail**', and (c) '**wide & long tail**' normalized to the referenced '**narrow**' group.

Analogous to **Fig, 5**, the effect of finite volume fraction can be clearly revealed by the ratio of $\gamma$ with respect to different volume fractions. As shown in **Fig. 7**, the relative value of $\gamma$ approaches faster to 1 with an increase in $\phi_V$. It means that increasing volume fractions can shorten the transient coarsening stage, which is consistent with of Chen and Voorhees's conclusions **[13]**. This finding is not so strange because of the increasing matrix diffusion by finite volume fractions, which can accelerate the many-body relaxation of diffusion-controlled transient coarsening. In addition, as compared by **Fig. 7(a-c)**, the differences between the three curves are larger for the wider distributions. In other words, the wider the initial distribution is, the more significant of increasing volume fractions for shortening the transient stage. This finding is also consistent with Li et al.'s phase-field simulation results **[17]**. Again, it shows that the '**supper narrow**'-like initial distribution can offer a coarsening system with a better thermal stability, consistent with the coarsening rate in **Fig. 3**.

### 5.3.2. Delaying effect on transient stage co-controlled by interface and matrix diffusion

Aimed at revealing the influence of finite volume fractions on transient coarsening co-controlled by interface and matrix diffusion, all simulations in this part apply the '**narrow**' as the initial distribution. The temporal evolutions of $R_{cr}$ against $t^{1/3}$ for $\phi_V = 0.1,\ 0.4,\ 0.7$ are plotted in **Fig. 8(a)**, where the long timescale provides a creditable examination for the effects of volume fractions. Using the final $\gamma$ values for **Case-1** with '**narrow**' PSD in **Section 4.2** as references $\gamma_s$, **Fig. 8(b)** exhibits the normalized values of $\gamma$ to characterize the proximity to steady stage.

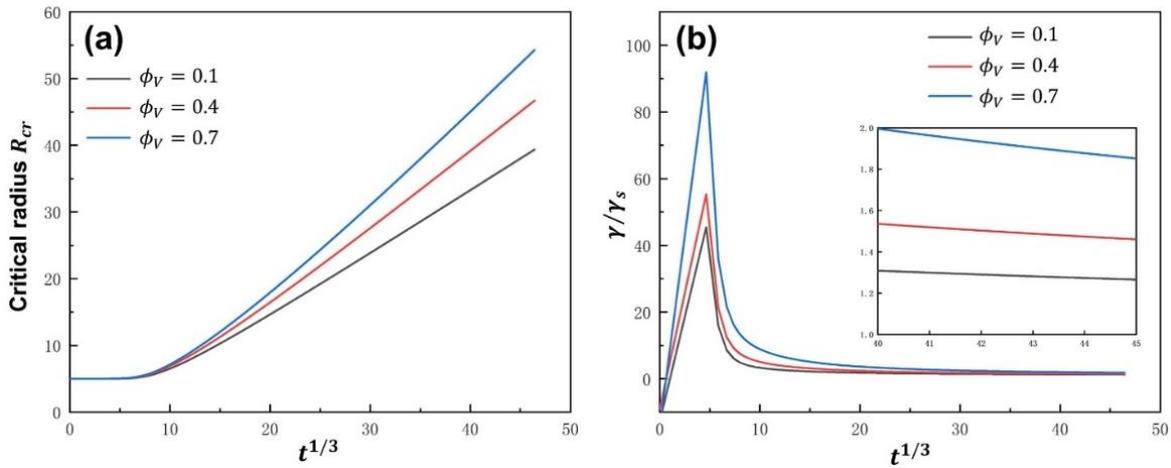

**Fig. 8.** (a) Temporal evolutions of $R_{cr}$; (b) $\gamma$ values for **Case-2** normalized to those of **Case-1**.

As compared in **Fig. 8(b)**, the ratio of $\gamma$ approach slower to 1 with an increase of volume fraction, which means that increasing volume fraction will extend the transient stage. This finding is opposite to the diffusion-controlled case in **Section 5.3.2**, but surprisingly, in agreement with Voorhees et al.' experimental observations [11]. As analyzed by White [34, 35] and Sun [36], among the interface migration (or reaction) and the matrix diffusion, the slower one will determine more for the coarsening kinetics. Due to the accelerating matrix diffusion by increasing volume fractions, the rate-determining strength of the interface process will be enhanced, thus, there will be a longer transient stage for the competition between two mechanisms. From a qualitative aspect, when the initial rate-determining strength of two mechanisms are comparable, the overall transient coarsening process may be divided into two substages: **(i)** from the initial state to the interface-dominated state, and **(ii)** from the interface-dominated state to the final diffusion-controlled state. Therefore, it is necessary to divide the transient coarsening stage and discuss the effects of volume fraction on these two substages, respectively.

In order to divide the substages, here is an **analytical approximation** for the second substage that evolved from interface-dominated state to diffusion-controlled state. For the case of $\phi_V \to 0$, White **[34, 35]** and Sun **[36]** have shown that the time-dependence of dimensionless distribution $h(\rho, \tau)$ is significantly smaller than that of the particle number density $g(\tau)$, and thus the left hand of **Eq. (30)** can be ignored like the steady stage. Under the similar scenario, we also solve the two parameters of **Eq. (25)** by the LSW-asymptotic analysis ($d\rho/d\tau = d^2\rho/d\rho d\tau = 0$) at an evolving $\rho_{\max}$, yielding as

$$\gamma_{2nd} = \frac{3Dd}{\Delta C_{eq}} \frac{dt}{dR_{cr}^3} = \frac{\rho_{\max}^3}{(2-\rho_{\max})(1+\beta_V \rho_{\max})^2}$$

$$\lambda_{2nd} = \frac{D}{KR_{cr}} = \frac{\beta_V \rho_{\max}^2 (2-\rho_{\max}) + \rho_{\max}(3-2\rho_{\max})}{(\rho_{\max}-2)(1+\beta_V \rho_{\max})^2}.$$
(36)

And the $h(\rho, \rho_{\max})$ can be numerically solved by the following equation

$$h(\rho, \rho_{\max}) = \exp\left\{\int_0^{\rho_m} -\left[\frac{\frac{d\ln g(\tau)}{d\tau} + \frac{d^2\rho}{d\rho d\tau}}{\frac{d\rho}{d\tau}}\right] d\rho\right\},$$
(37)

and the validity of this equation is proved by **Fig. 9**, where there is a good agreement with the present simulations for long numerical steps.

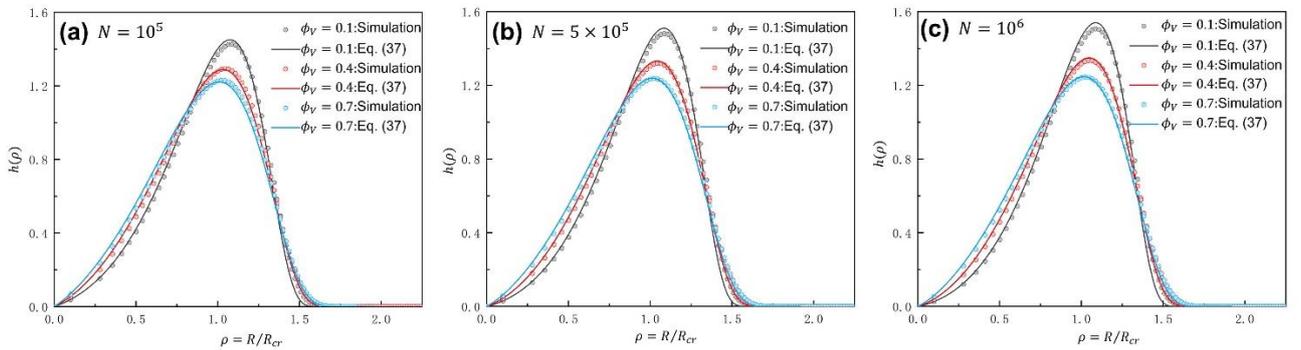

**Fig. 9.** Comparison of predictions between **Eq. (37)** and the numerical simulations.

Now, with a convincing analytical approximation of the second substage, these two substages can be separated by normalizing the simulated values $\gamma$ with the $\gamma_{2nd}$ in **Eq. (36)**. However, the present expression of $\gamma_{2nd}$ is inconvenient to directly applied because it is a function of $\rho_{\max}$. To simplify this problem, we here introduce an approximation for $\gamma_{2nd}$, as done by Sun [36]

$$\gamma_{2nd} \approx \frac{27}{4} \frac{(1-\phi_V^{1/2})(1+\phi_V^{1/2}/3)^3}{(1+\phi_V^{1/2})^2} + \frac{4D}{KR_{cr}}, \tag{38}$$

which can be viewed as a linear interpolation of the diffusion-controlled $\gamma$ [**Eq. (32)**] and Wagner's interface-controlled $\gamma$ [6]. As compared in **Fig. 10** for $\phi_V = 0.1, 0.4, 0.7$, there is an excellent agreement between **Eqs. (36)** and **(38)**, thus, the approximated form will be applied as a reference to separate the two substages. **Fig. 11** shows the ratio of simulated values normalized to **Eq. (38)**, from which it can be found that an increasing of volume fraction can accelerate the first substage that evolved from the initial state to the interface-dominated state. This is actually consistent with the former analysis in **Section 5.3.2**. Therefore, the dual effects of volume fraction are clarified. An increase in volume fraction can accelerate the first stage of transient coarsening, while delay the second stage.

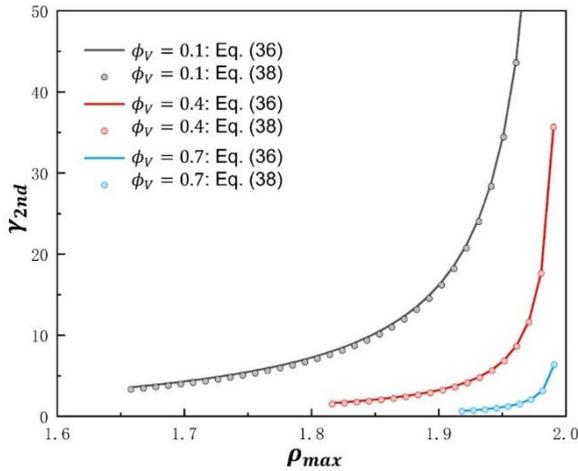 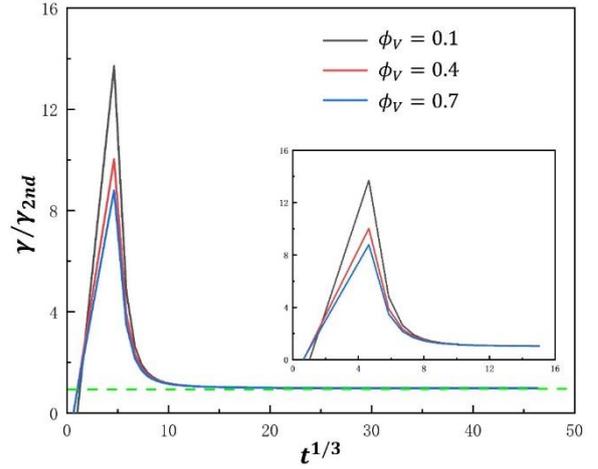

**Fig. 10.** Comparison between **Eqs. (36)** and **(38)**.    **Fig. 11.** $\gamma$ values for **Case-2** normalized to **Eq. (38)**.

## 5.4. Limiting coarsening behaviors at ultrahigh-volume fractions

In the previous section, it has been revealed that an increase in volume fraction can enhance the rate-determining strength of the interface process from a qualitative aspect. And this part will quantitatively analyze its influences on distributions and growth exponent of the second substage, which can also provide a promising explanation for Wang et al.'s phase-field simulations for coarsening at ultrahigh volume fraction regime **[37-40]**, where there is a descent of scaling growth exponent with an increase in volume fractions **[40]**.

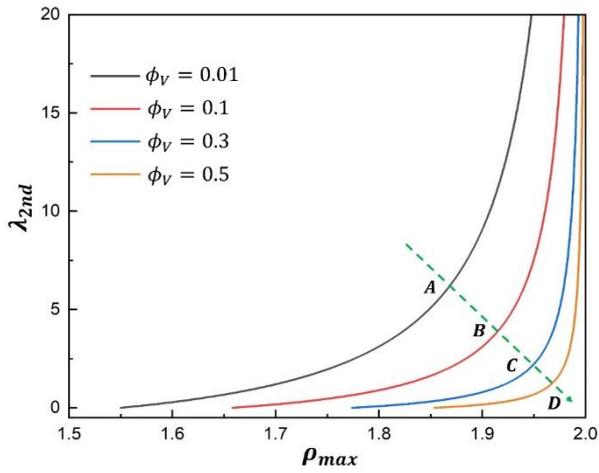
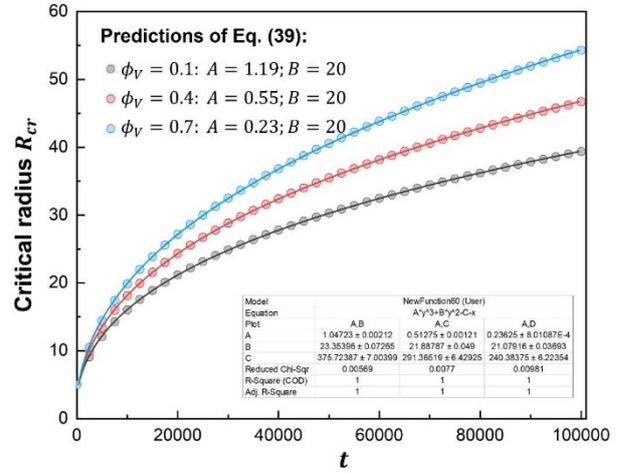

**Fig. 12.** Relationship between $\lambda_{2nd}$ and $\rho_{\max}$.   **Fig. 13.** Comparisons between **Eq. (39)** with simulations.

In the second substage evolved from the interface-dominated state to the diffusion-controlled state, the maximum relative radius $\rho_{\max}$ monotonically varies with $\lambda_{2nd}$, as given by **Eq. (36)**. And **Fig. 12** exhibits the relationship between $\lambda_{2nd}$ and $\rho_{\max}$ with different volume fractions. When $\lambda_{2nd} \to \infty$, the maximum relative radius $\rho_{\max} \to 2$ for the Wagner distribution [**Eq. (11)**]. When $\lambda_{2nd} \to \infty$, the maximum relative radius $\rho_{\max} \to 1.5 + 0.5\phi_V$ for the Streitenberger's distribution function [**Eq. (7)**]. In **Fig. 12**, there exists a series of the critical points, as marked by the green arrow ('**A-D**'). Only when $\lambda_{2nd}$ is lower than the critical point, the system can rapidly evolve to the diffusion-controlled distribution [**Eq. (7)**] and then reach the steady stage. In addition, the value of the critical point will significantly decrease with an increase in volume fraction, meaning that there will be a longer transient time of the second sub-stage. Therefore, it can be inferred that the observed distribution at high volume fraction regime will be close to the Wagner distribution, **Eq. (11) [6]** or Hillert distribution for normal grain growth **[63]**, which is consistent with Wang et al.'s simulation results **[37-40]**.

As for the growth equation of the critical radius $R_{cr}$, it can be derived from the $\gamma_{2nd}$ in **Eq. (38)**, yielding as

$$t + cons = \frac{9\Delta C_{eq}}{4Dd} \frac{(1-\phi_V^{1/2})(1+\phi_V^{1/2}/3)^3}{(1+\phi_V^{1/2})^2} R_{cr}^3 + \frac{2\Delta C_{eq}}{Kd} R_{cr}^2 \qquad (39)$$

and **Fig. 13** compares the predictions of **Eq. (39)** with those of the numerical simulations. Although the fitting range of the simulation results contains the first substage of transient coarsening, it is found that there still exists a good agreement between **Eq. (39)** and simulations. In addition, in order to make the effect of increasing volume fractions more transparent, **Eq. (39)** can be rewritten as

$$\begin{aligned} t + cons &= \frac{9\Delta C_{eq}}{4Dd} \frac{(1-\phi_V^{1/2})(1+\phi_V^{1/2}/3)^3}{(1+\phi_V^{1/2})^2} R_{cr}^3 (1+\eta); \\ \eta &= \frac{8d}{9\Delta C_{eq}} \frac{(1+\phi_V^{1/2})^2}{(1-\phi_V^{1/2})(1+\phi_V^{1/2}/3)^3} \frac{D}{KR_{cr}}. \end{aligned} \qquad (40)$$

When $\eta \to \infty$, **Eq. (40)** reproduces to the interface-controlled **Eq. (12)**. When $\eta \to 0$, the diffusion-controlled **Eq. (8)** can be reproduced. With an increase in critical radius, the scaling growth exponent will transfer from $2$ to $3$. As shown in **Fig. 14**, $\eta$ will increase along with the volume fraction $\phi_V$, especially for $\phi_V > 0.9$. The higher the $\eta$ is, the closer of growth exponent to $2$. At somewhat ultrahigh volume fractions, it is possible to observe an intermediate scaling growth exponent between $2$ and $3$, and the exponent will be approach to $2$ with increasing volume fractions. In means that the present theoretical analysis shows a good agreement with Wang et al.'s recent simulations and analysis **[40]**.

As a result, there exists a crossover from diffusion-controlled coarsening to interface-controlled coarsening at ultrahigh volume fractions, including both the shape of distribution and the scaling growth exponent. It is worthwhile noting that the present analysis did not consider the influence of the topological correlation at high volume fraction regime. When volume fraction is significantly high, the phase-coarsening microstructure is analogous to the grain-growth microstructure, where the shape of particle will be like a polyhedron **[71]**, and the larger particles will tend to be surrounded by the smaller one. In normal grain growth **[72]**, the local environmental heterogeneity can significantly affect the dynamic equation of individual grains and make distribution deviate from the Hillert distribution (without topological correlation) **[63]**. A similar influence is expected for phase coarsening at ultrahigh volume fractions, e.g., the simulated distributions by Wang and co-workers are much closer to the distribution function of some current authors' recent work for grain growth **[72]**, as shown in **Fig. 15**.

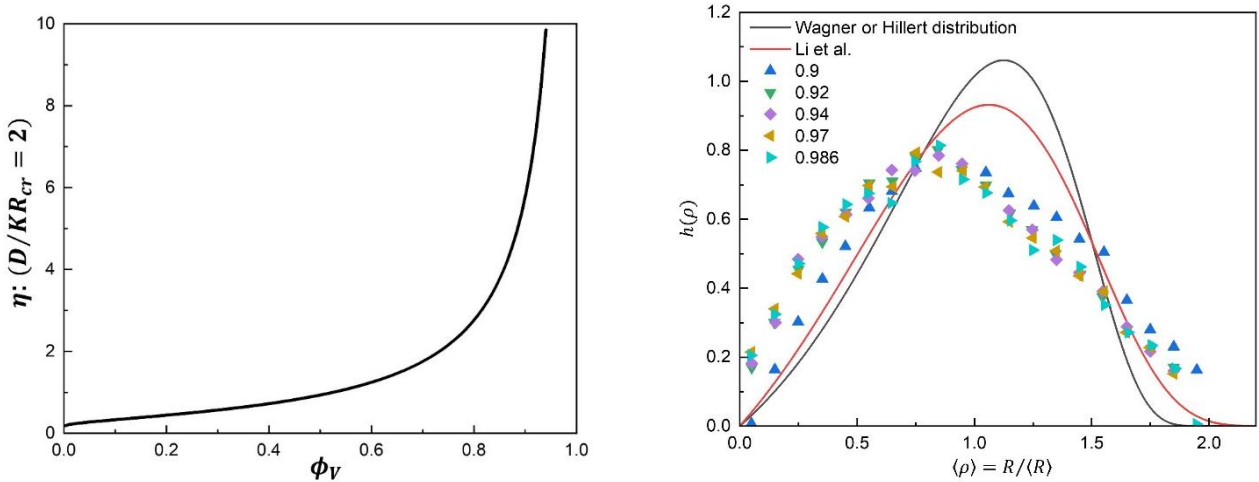

**Fig. 14.** Relationship between $\eta$ and $\phi_V$.  **Fig. 15.** Comparison of Wang et al.'s simulated distributions **[40]** with Wagner **[6]** or Hillert distribution **[63]** (without topological correlation) and Li et al.'s distribution **[72]** (with topological correlation).

## 6. Conclusions

Considering the free energy dissipations by matrix diffusion and interface migration, this work derived a general expression for phase coarsening kinetics. By introducing a dimension free and time-dependent distribution function $h(\rho,\tau)$, a highly efficient numerical framework in the dimensionless LSW space is established to revisit the effects of initial PSD and volume fractions on transient coarsening kinetics, which provided reasonable explanations for the existing debates about transient coarsening. The main conclusions are as follow:

(1) There is no monotonical relationship between the transient time length and width or tail length of the initial distributions. The ultralong transient stage, defined as 'quasi-steady' state, may occur for both the '**wide & long tail**' and '**supper narrow & short tail**', where the coarsening systems with narrower initial distribution can behave a better thermal stability.

(2) The final distributions at $10^6$ of the '**wide & long tail**' and '**supper narrow & short tail**' are consistent with the Brown's steady problem **[20-22]** in 1990s. In this article, the numerical origin of Brown's steady stage was firstly revealed, which is due to a 'meet by chance' between the instantaneous distribution $h(\rho,\tau)$ and a distribution $\bar{h}(\rho,\tau)$ (determined by instantaneous scaled evolution equation $d\rho/d\tau$, where the difference between $h(\rho,\tau)$ an $\bar{h}(\rho,\tau)$ determines the evolving rate of $h(\rho,\tau)$. In the further, more effects are still needed to reveal the evolving character of the transient PSDs.

(3) An increase in volume fraction can result in two exactly opposite effects on transient coarsening. On the one hand, it can shorten the transient time that evolved from the initial state to the final state (for the diffusion-controlled coarsening system) or from the initial state to the interface-dominated state (for the coarsening co-controlled by interface and matrix diffusion). On the other hand, it can enhance the rate-determining strength of the interface process, and thus delay the transient time that evolved from the interface-controlled state to the final diffusion-controlled state. This finding offers a promising explanation for Voorhees et al.'s experiment for transient coarsening **[11]**.

(4) At somewhat ultrahigh volume fraction regime, there is a crossover from diffusion-controlled phase coarsening to the interface-controlled coarsening, including both the shape of distribution and the scaling growth exponent, which qualitatively agreed with Wang et al.'s previous phase-field simulations for ultrahigh volume fraction regime **[37-40]**.

**Acknowledgments**

The work was supported by the Research Fund of the State Key Laboratory of Solidification Processing (NPU), China (Grant No. 2020-TS-06, 2021-TS-02), and Natural Science Basic Research Program of Shaanxi (Program No. 2022JC-28). Yue Li sincerely thanks Prof. Haifeng Wang in Northwestern Polytechnical University for lectures of the thermodynamic extremal principle.